\documentclass[aps,physrev,superscriptaddress,twocolumn,floatfix,10pt,showkeys,altaffillsymbol]{revtex4-2}

\usepackage{times}
\usepackage{amssymb,amsmath}
\usepackage{graphicx}
\usepackage{multirow}
\usepackage{float}
\usepackage{braket}
\usepackage{mathrsfs}
\usepackage{silence}
\usepackage{fancyvrb}
\usepackage{url}
\usepackage{hyperref}

\newcommand{\muB}{\mu_{\text{B}}}
\newcommand{\muN}{\mu_{\text{n}}}

\DeclareMathAlphabet{\mathbfit}{T1}{ptm}{b}{it}

\usepackage[svgnames]{xcolor}

\makeatletter
\newcommand{\fmarki}{*}
\newcommand{\fmarkii}{\ensuremath{\dagger}}
\newcommand{\fmarkiii}{\ensuremath{\ddagger}}
                
\def\@fnsymbol#1{{\ifcase#1\or \fmarki\or \fmarkii\or \fmarkiii\or \fmarkiv\or \fmarkv\or \fmarkvi\or \fmarkvii\or \fmarkviii\or \fmarkix \else\@ctrerr\fi}}
\makeatother

\renewcommand{\fmarki}{*}
\renewcommand{\fmarkiii}{*}

\renewcommand{\appendixname}{Supporting Information}

\begin{document}

\title{Transient Plastic Spin Labeling with Chlorine Dioxide}

\author{Bence~G.~M\'{a}rkus$^\dagger$}\email[Corresponding author: ]{bmarkus@nd.edu}
\affiliation{{Stavropoulos Center for Complex Quantum Matter, Department of Physics and Astronomy, University of Notre Dame, Notre Dame, Indiana 46556, USA}}

\author{S\'andor~Kollarics}
\thanks{These authors contributed equally.}
\affiliation{{Department of Physics, Institute of Physics, Budapest University of Technology and Economics, M\H{u}egyetem rkp. 3., H-1111 Budapest, Hungary}}
\affiliation{{Institute for Solid State Physics and Optics, HUN-REN Wigner Research Centre for Physics, PO. Box 49, H-1525, Hungary}}

\author{Krist\'of~K\'aly-Kullai}
\affiliation{{Department of Physics, Institute of Physics, Budapest University of Technology and Economics, M\H{u}egyetem rkp. 3., H-1111 Budapest, Hungary}}

\author{Bernadett~Juh\'asz}
\affiliation{{Department of Physics, Institute of Physics, Budapest University of Technology and Economics, M\H{u}egyetem rkp. 3., H-1111 Budapest, Hungary}}

\author{D\'avid~Beke}
\affiliation{{Institute for Solid State Physics and Optics, HUN-REN Wigner Research Centre for Physics, PO. Box 49, H-1525, Hungary}}
\affiliation{Kand\'o K\'alm\'an Faculty of Electrical Engineering, \'Obuda University, Tavaszmez\H{o} u. 17., H-1084 Budapest, Hungary}

\author{L\'aszl\'o~Forr\'o}
\affiliation{{Stavropoulos Center for Complex Quantum Matter, Department of Physics and Astronomy, University of Notre Dame, Notre Dame, Indiana 46556, USA}}

\author{Zolt\'an~Noszticzius}
\affiliation{{Department of Physics, Institute of Physics, Budapest University of Technology and Economics, M\H{u}egyetem rkp. 3., H-1111 Budapest, Hungary}}

\author{Ferenc~Simon}\email[Corresponding author: ]{simon.ferenc@ttk.bme.hu}
\affiliation{{Department of Physics, Institute of Physics, Budapest University of Technology and Economics, M\H{u}egyetem rkp. 3., H-1111 Budapest, Hungary}}
\affiliation{{Institute for Solid State Physics and Optics, HUN-REN Wigner Research Centre for Physics, PO. Box 49, H-1525, Hungary}}

\date{\today}

\begin{abstract}\centering
    \includegraphics[width=0.6\columnwidth]{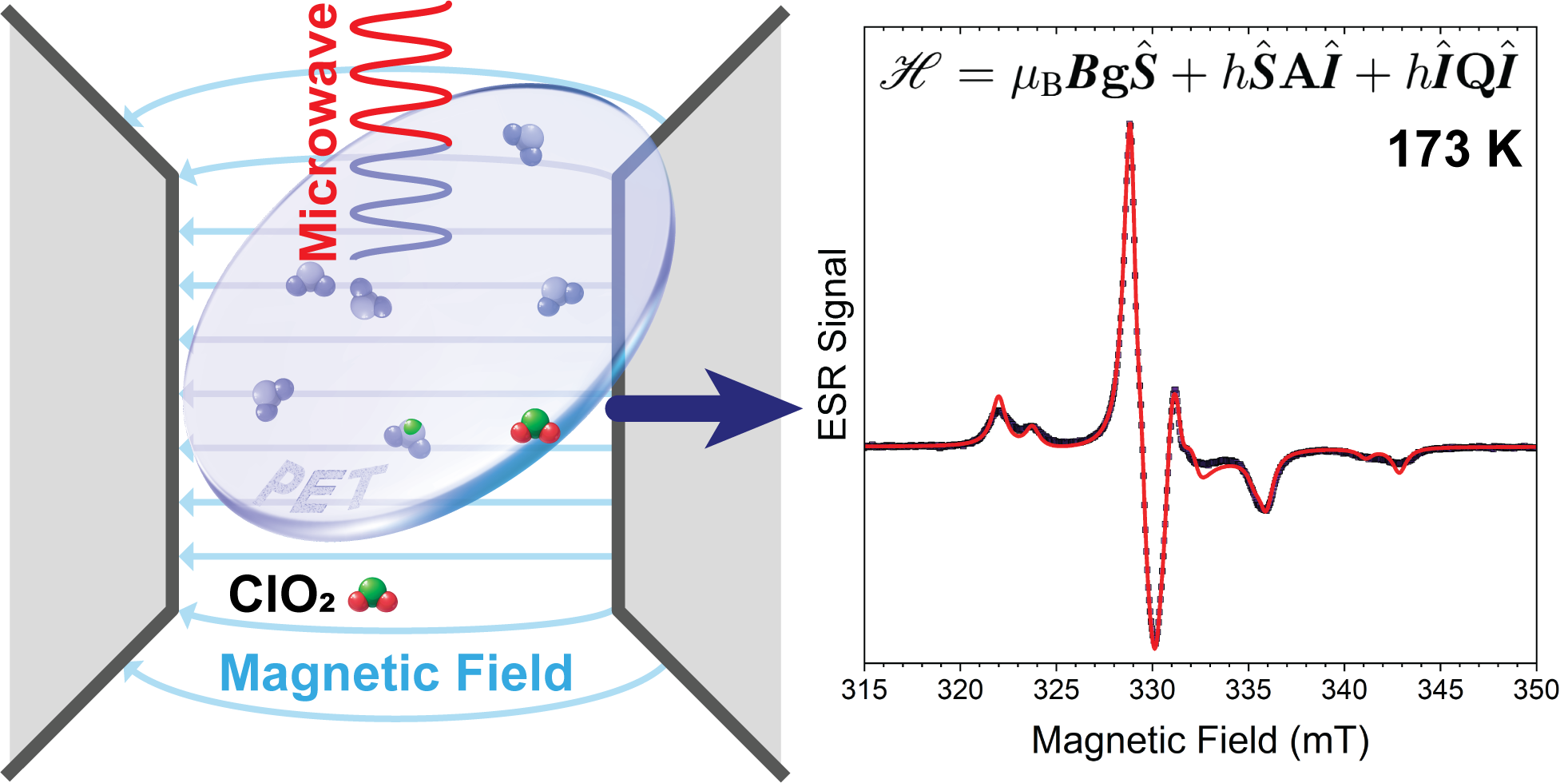}\\
    \textbf{ABSTRACT:} Plastic waste, being one of the most important problems for humankind, poses severe threats to ecosystems, wildlife, and human health. Tracing, quantifying, and identifying types of plastic waste is of crucial importance to understand its environmental pathways and develop targeted strategies for reduction, recycling, and remediation. To contribute to addressing this global issue, we investigated the spin-labeling capabilities of aqueous chlorine dioxide (ClO$_2$) radicals upon introduction into poly(ethylene terephthalate) and utilized electron spin resonance spectroscopy for detection. The technique is capable of identifying plastic species as the unpaired electron of the radical molecule is strongly sensitive to its local environment through its coupling parameters. Temperature-dependent measurements revealed that the molecules are immobilized at low temperatures and exhibit well-resolved anisotropic and hyperfine spectra that are quantitatively described by a model spin Hamiltonian. Even above the melting point of water, some degrees of freedom remain blocked as a result of the polymer matrix. Furthermore, employing a time-series measurement at room temperature enabled us to determine the diffusion coefficient of the molecule in the polymer.
\end{abstract}


\maketitle

\section{Introduction}

The global accumulation of plastic waste represents one of the most pressing contemporary environmental challenges \cite{Jambeck2015Science, Geyer2017SciAdv}. Despite increasing efforts in recycling and waste management, the long-term stability and chemical inertness of common polymers such as polyethylene terephthalate (PET) hinder their degradation and traceability in the environment \cite{Evode2021CSCEE, Zhang2021JCP, Pilapitiya2024CM}. Microplastics are even posing a more severe concern due to their health and biological effects and potential hazards \cite{Rochman2018Science, Hale2020JGRO}. Consequently, the reliable analytical characterization of polymeric materials has become a major scientific and technological challenge, driving extensive research efforts to address this pressing issue~\cite{vasudeva2025Trac, matavos2024EA, kumara2024Trac, lv2021WER, baruah2022EST}. The development of molecular-scale probes capable of detecting, identifying, and quantifying polymeric materials is, therefore, of great scientific and technological interest.

Current detection methods rely mainly on microscopy combined with infrared or Raman spectroscopy, and on thermal analysis, which, although chemically informative, are limited by low throughput, matrix interferences, and fluorescence~\cite{baruah2022EST}.

In this context, approaches that enable non-optical detection and quantification of polymers are particularly desirable. Spin labeling provides a complementary strategy by enabling direct detection of plastics through paramagnetic probes, and is widely used in biological systems, where nitroxide radicals have become standard reporters of local structure and motion \cite{Griffith1969ACR, Raikov2001MH, Berliner2002book, Haugland2018CSR, Torricella2021PP}. However, extending such methodologies to synthetic polymers remains challenging, primarily due to difficulties in incorporating stable radicals into inert matrices and maintaining their integrity over extended timescales \cite{Yamada1999PPS, Lucarini2003PSS, Krinichnyi2014APR, Uddin2020JPS}.

Chlorine dioxide (ClO$_2$) is a small and stable paramagnetic radical and a powerful oxidant \cite{Gan2020ESWR}. Due to its antibacterial and antiviral properties, it is widely used to treat drinking water and employed as an oral disinfectant \cite{Huang1997WR, Junli1997WR, Chauret2001AEM, Noszticzius2013PLoSOne}. Since the ClO$_2$ molecule has an unpaired electron residing in a $\pi^{\ast}$-antibonding orbital, it can be detected using electron spin resonance (ESR) spectroscopy \cite{Bennett1956PPSA, Pietrzak1970JCP, McDowell1973JCP, Weil2007book, Eaton2010book}. Despite the free radical nature of the molecule, since the electron on the $\pi^{\ast}$-orbital is participating in both oxygen bonds, this resonance hybrid is stabilizing the system and is responsible for its relative inertness \cite{Pauling1988GenChem, Flesch2006JMS}. Therefore, unlike typical organic spin labels, ClO$_2$ can diffuse into polymers without reacting with the host matrix, thus providing a simple route to \emph{in situ} spin labeling.

Indeed, ESR spectroscopy provides an effective approach for characterizing the local environment and dynamics of the spin label molecules in solids and polymers \cite{McConnell1970QRB, Tormala1979JMS, Savitsky2008JPCB, Xia2011JACS, Uddin2020JPS}. Furthermore, spin labeling techniques are widely employed in biological systems, where nitroxide radicals have become standard reporters of local structure and motion \cite{Griffith1969ACR, Raikov2001MH, Berliner2002book, Haugland2018CSR, Torricella2021PP}. ESR is also known to have superior sensitivity to detect free radicals in the bulk, such as ClO$_2$. Given the $1\times10^9$ spins/G sensitivity of the instrument (as specified by the manufacturer), a quick and conservative estimate yields a limit of detection of $<0.1$~ppm for ClO$_2$. Here, the volume of the solution was estimated to be around $100~\mu$L, and the spectral linewidth to be $2$~mT. A factor of $1{,}000$ was also considered in the molarity to account for the high dielectric losses caused by the water medium \cite{NIH2025}. This is also confirmed by the signal-to-noise ratio of our spectrum for the $30$~ppm concentration.

In addition to structural insights, spin labeling can provide a route toward the molecular tagging and tracing of plastic materials. The proposed method could complement recently developed fluorescent plastic labeling methods \cite{Gao2017ASS, Woidasky2020RCR, Grisi2024M, Moon2024SciAdv, Xie2025ACSSRM, Zhang2026CRPS}. Such an approach could facilitate the monitoring of degradation processes or the identification of polymer types in mixed waste streams. For this purpose, radicals that can penetrate and remain stable within dense polymer matrices are particularly desirable.

In this work, we demonstrate that ClO$_2$ radicals can be efficiently introduced into PET, where they remain stable over a wide temperature range and for sufficiently long times. Temperature-dependent ESR measurements reveal hindered rotational dynamics consistent with confinement within the voids of the polymer strings. The experimental data are accurately reproduced by a modeled spin Hamiltonian. Time-dependent measurements furthermore allowed the determination of the diffusion coefficient in the material.

Our findings establish chlorine dioxide as an effective and chemically simple spin label for polymeric materials, opening new perspectives for the ESR-based study and the tracing of plastics.

\section{Methods}

\paragraph{Materials}

High-purity aqueous solution of chlorine dioxide (ClO$_2$) was prepared by \textit{in situ} generation from sodium chlorite, followed by selective permeation of the ClO$_2$ gas through a nonporous polymer membrane made of silicone rubber into distilled water. This membrane-permeation method, originally developed by Noszticzius and co-workers \cite{NZpatent}, yields a hyperpure ClO$_2$ solution, meaning that the solution is not contaminated with reagents or by-products, because they cannot permeate through the membrane. Such hyperpure solutions are even suitable for biological and physicochemical applications \cite{Noszticzius2013PLoSOne, Noszticzius2013arXiv, KalyKullai2020, KalyKullai2025}.

The ClO$_2$ concentration in the collected aqueous phase is usually between $3000$ and $3500$ ppm. Its exact value was determined by iodometric titration \cite{Noszticzius2013PLoSOne}, and it was diluted by distilled water according to the required concentration. All solutions were kept cold, protected from light, and in a sealed bottle to minimize volatilization and decomposition.

Various concentrations of aqueous solutions of ClO$_2$, ranging from $30$ to $3000$ ppm, were prepared and transferred to a $10$~mL brown glass vial. A PET film with a lateral size of $1~\text{cm} \times 4~\text{cm}$ and a thickness of either $12$ or $100~\mu$m was immersed in the solution. The vial was sealed and stored in a refrigerator for at least $3$ days to achieve steady state, during which ClO$_2$ diffused homogeneously throughout the polymer matrix. The film was removed from the solution immediately before the ESR experiments, and any residual liquid was gently wiped off. A small piece (approximately $5~\text{mm} \times 5~\text{mm}$) was then cut and placed into a quartz tube with an internal diameter of $4~\text{mm}$. When time-dependent measurements were performed, a continuous dry nitrogen flow was applied to remove the chlorine dioxide molecules diffusing out of the PET film. This technique allowed us to determine the diffusion coefficient in PET from the decrease in the ESR intensity.

\paragraph{Electron Spin Resonance}

Electron spin resonance (ESR) measurements were carried out using an X-band ($\sim 9.4$ GHz) spectrometer operating in continuous-wave mode. The samples were placed in standard quartz tubes with an internal diameter of $4$~mm and cooled or heated as required using a variable-temperature unit. Time-dependent measurements were performed in a quartz tube with open ends, and a continuous dry nitrogen flow was applied to remove the chlorine dioxide molecules that diffused out of the PET film. This technique allowed us to determine the diffusion coefficient in PET from the decrease in the ESR intensity. Magnetic field modulation and lock-in detection were employed to record the first derivative of the absorption signal. Here, we wish to note that while the temperature-dependent measurements were acquired in a variable quartz cryostat, the time-dependent measurements were performed without this extension. The dielectric nature of the cryogenic quartz insert lowers the resonance frequency by about $400$ MHz (and thus the resonant magnetic field by about $14$ mT). These differences are apparent from a comparison of Figs. \ref{Fig2} and \ref{Fig3:diffusion}. The $g$-factors and linewidths were determined from the field positions and shapes of the resonance lines and following the same procedures as in our previous studies \cite{Markus2020ACSNano, Markus2023NatCommun}. Calibration of the magnetic field and microwave frequency was performed using a $1.5$~ppm Mn:MgO reference with a known $g$-value of $2.0014$. The modulation of the magnetic field was $0.1$~mT for the concentration and temperature-dependent measurements. For the time-dependent measurements, a higher, $0.5$~mT modulation was employed to enable faster accumulation times. No significant distortion of the ESR lines was observed. A microwave power of $0.2$~mW and $0.02$~mW were used for the concentration-dependent and temperature-dependent measurements of the pure solutions, respectively. For the $100~\mu$m PET film, a lower microwave power of $2~\mu$W was employed. Time-dependent measurements on the $12~\mu$m PET films were recorded using $2$ mW power. No saturation effects were observed during the measurements. The ESR linewidth is defined as the half-width at half-maximum, $\Delta B = \text{HWHM}$ of the spectrum.

\paragraph{Modeling}

Theoretical modeling of the ESR spectra was performed using the EasySpin 6 software package \cite{Stoll2006JMR} using Matlab R2024b. Least squares fitting of parameters was achieved using the Nelder--Mead simplex algorithm, with all fits converged to the tolerance for the error function of $10^{-9}$. For the aqueous solution, the \verb+garlic+, for the solid spectra, the \verb+pepper+ functions were used. Above $273$ K for the PET samples, the \verb+chili+ function provided the best results. According to the documentation, the \verb+garlic+ function is for isotropic and fast-motional cw EPR spectra of radicals in solution. The \verb+pepper+ is for solid-state cw EPR spectra for powders, films, and crystals. Lastly, the \verb+chili+ function describes systems in between the previous two, tumbling spin systems in the slow-motional regime \cite{Stoll2006JMR}. The details of the spin Hamiltonian used are discussed in Section \ref{sec:res}. Relative orientations of the tensors in the molecular frame were neglected during the calculations.

\section{Results and Discussion}\label{sec:res}

Without any restrictions, an $S=1/2$ electronic spin interacting with an $I>1/2$ nuclear spin in a finite $\mathbfit{B}$ external magnetic field can be described with the following Hamiltonian:
\begin{equation}
    \mathscr{H} = \muB \mathbfit{B} \mathbf{g} \hat{\mathbfit{S}} + h \hat{\mathbfit{S}} \mathbf{A} \hat{\mathbfit{I}} + h \hat{\mathbfit{I}} \mathbf{Q} \hat{\mathbfit{I}} -\muN g_{\text{n}}\mathbfit{B}\hat{\mathbfit{I}},
\end{equation}
where the first term is the regular electronic Zeeman interaction between the electron spin, $\hat{\mathbfit{S}}$, and the external magnetic field, $\mathbfit{B}$, the second is the hyperfine term, describing the interaction between the electron and the nuclear spin, $\hat{\mathbfit{I}}$, and the third term is arising from nuclear quadrupolar effects (or the electric field gradient generated by the nucleus). The last term is the nuclear Zeeman term describing the interaction between the nucleus and the magnetic field, whose effect is negligible for ESR experiments. The coupling constants, the $g$-factor, $\mathbf{g}$, the hyperfine coupling, $\mathbf{A}$, and the quadrupole coupling, $\mathbf{Q}$, are rank-three tensors. The Bohr and nuclear magnetons are denoted with $\muB$ and $\muN$, respectively, and $h$ is the Planck constant.

Since the quadrupole tensor is traceless (e.g., $Q_\text{x}+Q_\text{y}=-Q_\text{z}$), it is common to introduce nuclear quadrupole coupling constant as $e^2Qq/h=2I(2I-1)Q_\text{z}$, and the $\eta=(Q_\text{x}-Q_\text{y})/Q_\text{z}$ asymmetry parameter, which can take values between $0$ and $1$. Here, $e$ is the elementary charge, $q$ is the largest component of the electric field gradient tensor at the nucleus, and $Q$ is the electric quadrupole moment of the nucleus.

\begin{figure}[htp]
    \includegraphics[width=\columnwidth]{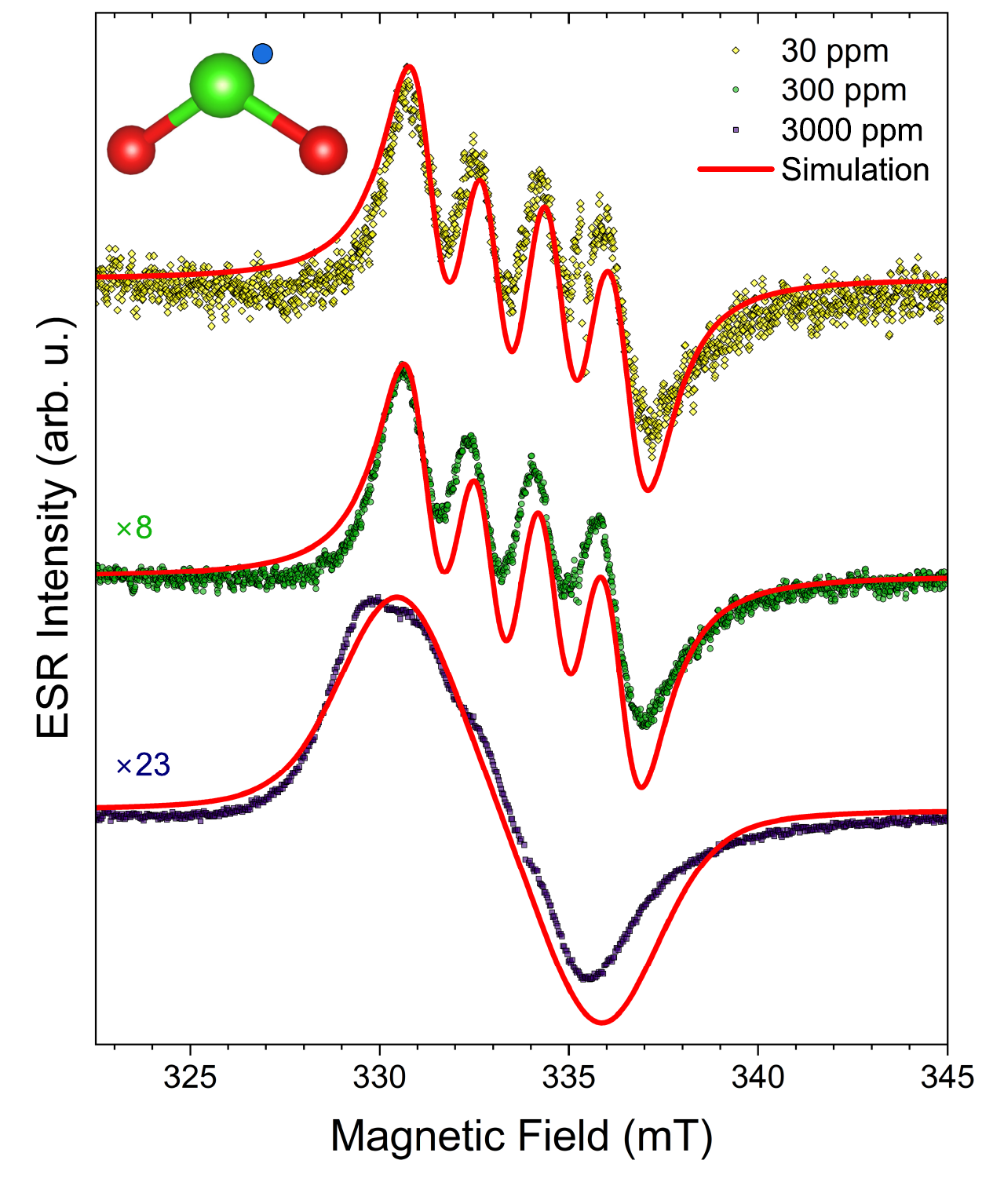}
    \caption{ESR spectra of ClO$_2$ in aqueous solution with three different concentrations, $30$, $300$, and $3000$ ppm at $298$ K. Multipliers indicate the relative intensity increase, determined by double integration, compared to the $30$ ppm solution. While a tenfold increase in concentration from $30$ to $300$ ppm increases the signal almost tenfold, above this threshold, the solution becomes too concentrated and broadening effects occur due to stronger interactions between the molecules. Solid curves are simulations from the spin-Hamiltonian with parameters fitted to match the experimental data. A detailed analysis is found in the text.}
    \label{Fig1:Cspectra} 
\end{figure}

\begin{figure*}[htp]
    \includegraphics[width=.98\textwidth]{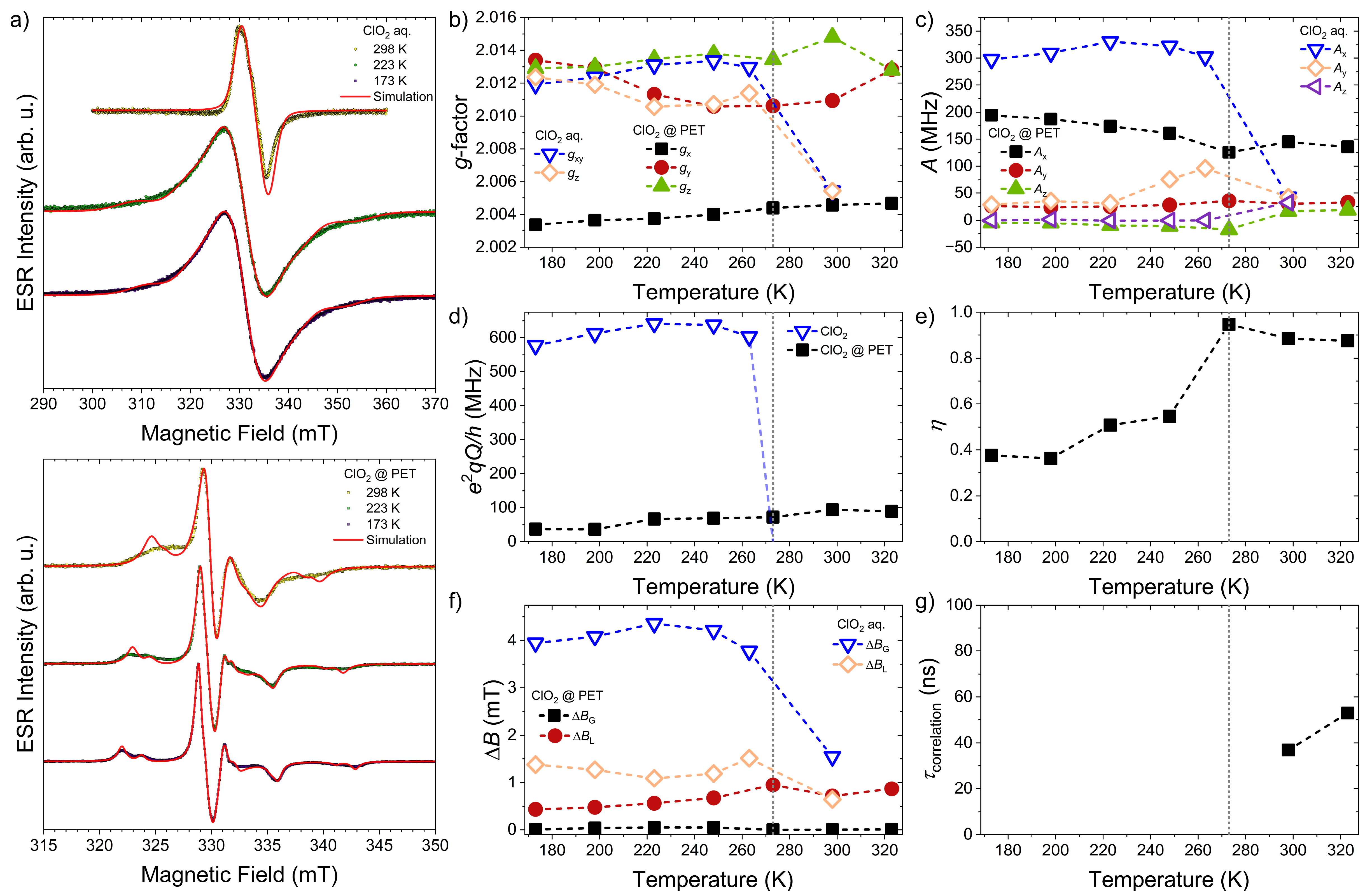}
    \caption{a) Detected derivative cw ESR spectra of ClO$_2$ in a $3000$~ppm aqueous solution at $298$, $223$, and at $173$ K (top) and the same for $100~\mu$m thick PET sample soaked in the solution for days. Please note the drastic change in the lineshape depending on the solid matrix. Solid lines are simulations with parameters given in panels b to g. The details of the spin model can be found in the main text. b-g) Parameters obtained from the spin-model: $g$-factors, hyperfine couplings ($A$), nuclear quadrupole coupling constant ($e^2qQ/h$) and its asymmetry parameter ($\eta$), linewidths ($\Delta B$) and rotational correlation time ($\tau_{\text{correlation}}$), respectively. In the solid phase, all parameters show little variation with temperature. On the other hand, around the phase transition of water, they do change drastically. Moreover, please note the significant difference between the aqueous solution and when the molecules are embedded in the PET matrix. Vertical dashed lines denote the melting point of water under ambient pressure. The semi-transparent dashed blue line in panel d suggests that the quadrupolar coupling averages to zero at $273$ K in the aqueous ClO$_2$.}
    \label{Fig2} 
\end{figure*}

Considering the case of ClO$_2$, there is a single unpaired electron with an $S=1/2$ spin, which interacts with the chlorine atom with a nuclear spin of $I=3/2$. Here, care was taken that Cl has two stable isotopes with a natural abundance of $75.8\%$ $^{35}$Cl and $24.2\%$ $^{37}$Cl with the same nuclear spin. The effect of the two oxygens is neglected as only $^{17}$O has a finite nuclear spin of $I=5/2$, but its abundance is only $0.0367\%$. Considering the most concentrated sample of $3000$ ppm investigated, this would mean a relative $^{17}$O concentration as little as ${\sim}1$ ppm. Interactions with the protons in water and ice are averaged out and only considered through the linewidth. The number of expected peaks to appear in an aqueous solution is thus $N=2I+1=4$ for each Cl isotope. Here, we note that due to the interaction with the solvent media, the observed ESR spectra are strongly solvent and environment dependent \cite{Bennett1956PPSA, Adams1966JCP, Vanderkooi1966IC}.

The ESR spectra of ClO$_2$ in aqueous solution with three different concentrations, $30$, $300$, and $3000$ ppm are shown in Fig. \ref{Fig1:Cspectra}. Low and moderate concentrations of $30$ and $300$ ppm allows the hyperfine peaks arising from the chlorine atom to be resolved; however, at $3000$ ppm a broadening effect prohibits to observe the four distinct peaks. The observed intensity should be linear with concentration for highly diluted solutions. This effect is observed for the $30$ ppm to $300$ ppm dilutions, where the tenfold concentration increase results in an almost tenfold (eight) intensity increase. Yet, above this concentration, the dipole interaction between the molecules becomes significant, and the signal broadens, prohibiting the resolution of the individual hyperfine peaks. This means that even after calibration, ESR determination of the concentration is only valid up to ${\sim}300$ ppm, as the observed intensity becomes non-linear above that in the aqueous solution. Here, intensities are determined by the double integration of the spectra.

In the liquid state for the two smaller concentrations, we observe an isotropic $g$-factor of around $2.0093(3)$ expected from the averaging due to fast motion of the molecules and in a relatively good agreement with the findings of Bennett \emph{et al}.\cite{Bennett1956PPSA} with $g=2.010 \pm 0.003$, and Ozawa \emph{et al}.\cite{Ozawa1983CPB, Ozawa1996FRBM} with ${g=2.0106}$. Interestingly, at the highest concentration we observed a lower $g$-factor value of $g=2.0054(3)$.

For the hyperfine interaction, we found that a uniaxial coupling of $A_\perp=65(1)$ MHz and $A_\parallel=19(2)$ MHz for the $30$ and $300$ ppm dilutions describes the system the best (in terms of reduced $\chi^2$ values). At low temperatures, in the solid phase, rotational and motional degrees of freedom are completely frozen out, and therefore, only powder averaging occurs. On the other hand, in a solution, motional averaging occurs: the $\Delta \omega$ static spectral width of the frozen solution is motionally averaged by the rotational correlation time, $\tau$, when $\Delta \omega \tau \ll 1$. This usually results in a spectrum showing equidistantly split hyperfine lines (the splitting gives the isotropic hyperfine constant). However, the spectrum is very complicated for intermediate values of $\Delta \omega \tau$, resulting in a spectrum that is neither of a perfect powder distribution nor of a perfect motionally averaged type. We unfortunately have no additional information about the expected molecular rotational correlation time of ClO$_2$; we thus cannot independently verify which regime is encountered herein. At the same time, Bennett \emph{et al}. \cite{Bennett1956PPSA} found an isotropic coupling of $A=48$ MHz (albeit their spectral resolution was limited at that time), which can be interpreted as a weighted average of our values, $A_{\text{iso}}=(2A_\perp+A_\parallel)/3\approx50(1)$ MHz. Using an isotropic model, we obtained $A_{\text{iso}}=49(1)$ MHz, in agreement with the averaged and the literature value. Oddly, at $3000$ ppm concentration, the observed hyperfine coupling becomes nearly isotropic with values of $A_\perp=42(1)$ MHz and ${A_\parallel=32(2)}$ MHz (or $39(1)$ MHz in the isotropic model).

\begin{figure*}[htp]
    \includegraphics[width=\textwidth]{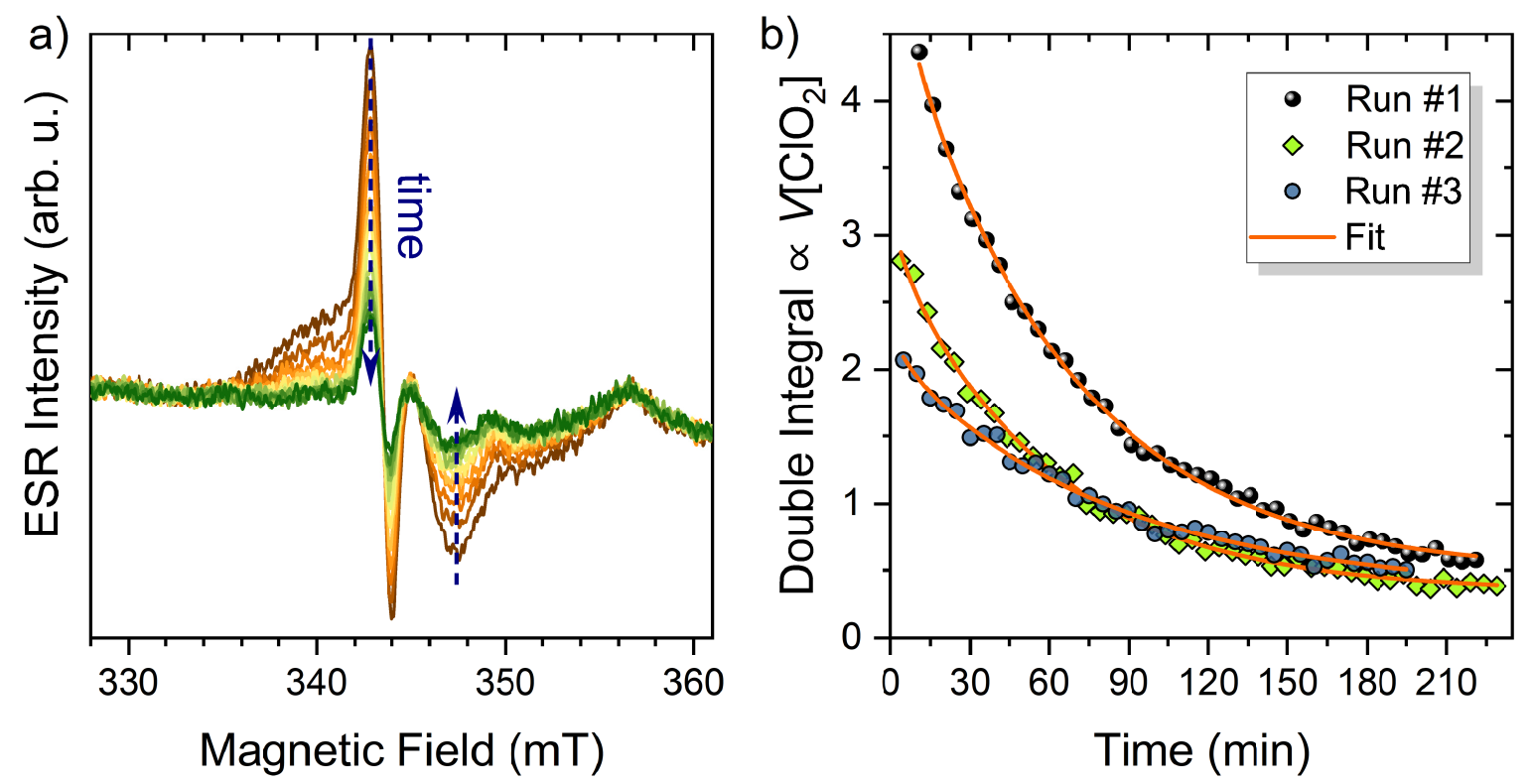}
    \caption{a) A time series of ESR measurements performed on a $12~\mu$m thick PET film, initially soaked in a $2900$ ppm aqueous solution of ClO$_2$. All measurements were performed at room temperature. The feature at $357$ mT arises from the sample holder and was excluded from later evaluation steps. The time difference between two adjacent spectra is $5$ minutes. Please note the systematic decrease in the spectral intensity. b) Time dependence of the ESR intensity from double integration of the spectrum for three separate runs. The curves follow an exponential decay in time as dictated by the out-diffusion and removal of the ClO$_2$ molecules from the PET film. The orange curve is a fit based on the theoretical discussion presented in the main text. Please note that the decay characteristic is independent of the initial settings. The obtained diffusion coefficient is $D=(3.91 \pm 0.74)\times 10^{-15}$~m$^2$/s at room temperature.}
    \label{Fig3:diffusion} 
\end{figure*}

Upon cooling down, below the freezing point of water, a considerable broadening of the ESR line is observed as shown in Fig. \ref{Fig2}a and f. This is in contrast to other -- organic \cite{Adams1966JCP} and inorganic \cite{Vanderkooi1966IC} -- solvents, where cooling narrows the line. Here, this is only observed below $220$ K. The line is also inhomogeneously broadened, which is considered through a Gaussian component of the linewidth. In the solid-state, most of the parameters differ from the ones observed in the liquid phase, as interactions are not averaged out anymore. In water-ice, the $g$-factor becomes uniaxial, while the hyperfine coupling has three distinct directions, as presented in Fig. \ref{Fig2}b and c, respectively. A strong, but isotropic quadrupolar coupling is also observed, in agreement with previous measurements on ClO$_2$ trapped in inert matrices at low temperatures \cite{McDowell1973JCP}. Nonetheless, most spectral parameters show little temperature dependence in the observed temperature region. Notably, the anisotropies of the $g$-factor, the hyperfine coupling, and the quadrupole coupling have a weak maximum between $220-250$ K.

When introduced into PET, all spectral observables become rhombic for both below and above the freezing point of water. This indicates that even in a liquid solution state, the polymer matrix prevents certain motions and interactions from being averaged out partially. A similar behavior is observed in KClO$_4$ \cite{Cole1960PNAS}, synthetic zeolites \cite{Pietrzak1970JCP}, and on the surface of MgO \cite{Tench1971JCS}, where the crystal symmetry strongly affects the coupling constants. Compared to the water-ice matrix, the anisotropy of the $g$-factor is much more pronounced for the $g_\text{x}$ direction (Fig. \ref{Fig2}b). The magnitude of the hyperfine coupling, as well as its asymmetry, is moderately reduced; in turn, the quadrupole coupling is reduced by an order of magnitude, as shown in Figs. \ref{Fig2}c and d. The latter also displays a strong anisotropy with $\eta$ values close to $1$ above the freezing point and around $0.5$ below (Fig. \ref{Fig2}e). Furthermore, the observed linewidth is significantly narrower and almost no Gaussian broadening is observed (Fig. \ref{Fig2}f). Most parameters, except the anisotropy of the quadrupole coupling, have little temperature dependence again. Above the melting point of water, neither the solid nor the fast-moving liquid model describes the system adequately. Instead, a slowly-moving, tumbling description was found to be appropriate, where a rotational correlation parameter was introduced and is in the order of $50$ ns as shown in Fig. \ref{Fig2}g. Another intriguing observation is that, even though the observed intensity as a function of concentration in the aqueous solution starts to deviate from a linear relation above $300$~ppm, the PET matrix seems to alleviate this limitation as the ClO$_2$ molecules are more separated, allowing reliable intensity determination in PET films soaked in up to $3000$~ppm aqueous solutions of ClO$_2$.

The fit quality of ClO$_2$ embedded in the PET is reasonably good at low temperatures (e.g., at $173$ K); however, it is apparently inferior around room temperature. This is most probably due to an ill-defined rotational axis and rotational correlation time, as well as the possible presence of a distribution in the latter. This may arise from the non-uniform spacing/voids of the PET matrix. At low temperatures, the fully static, randomly oriented powder distribution of the molecules is a well-defined physical state. In contrast, a partially hindered rotation with a distributed correlation time cannot be uniquely simulated.

Lastly, we study the kinetics of how ClO$_2$ diffuses out from the PET. The previously soaked samples were placed in a continuous flow of dry nitrogen gas to remove the outgassed molecules, and a time-series of ESR spectra was accumulated as presented in Fig. \ref{Fig3:diffusion}a. Accumulation of a single spectrum took about $42$ seconds, and the time difference between the accumulations was $5$ minutes. Here, it has to be noted that preparation of the measurement required some time (e.g., transferring the sample and adjusting the gas flow). This time was measured and is considered a finite shift in the time axis.

The decrease in the ESR signal can be modeled as a diffusion process, as the ClO$_2$ molecules leave the PET film. Using Fick's laws, the diffusion equation for the $c(\mathbfit{r},t)$ concentration in the film reads
\begin{equation}
    \frac{\partial c(\mathbfit{r}, t)}{\partial t} = D\nabla^2 c(\mathbfit{r},t),
\end{equation}
where $D$ is the diffusion coefficient, assumed to have no spatial dependence. The obtained ESR intensity is directly proportional to the total amount of substance -- and hence the integral of the concentration -- as there are no other spin species present in the system, ${I_{\text{ESR}} \propto \int_Vc = V[\text{ClO}_2]}$. Denoting the surface of the cross-section of the film with $A$, and the thickness with $L$, the bulk volume, measured by the ESR, is $V=A\times L$. As we have no spatial resolution, the concentration has to be integrated for the whole bulk for every $t$. Since the thickness of the film is considerably smaller, most molecules will leave the film through the faces with $A$ area. This simplifies the mathematical problem to a single spatial dimension, $x$, which is taken along the thickness of the film. This also means that iso-concentration surfaces are planes parallel to the $A$ area. Before the measurement, the samples were soaked in the ClO$_2$ solution sufficiently long to reach a steady state, meaning that the initial concentration of ClO$_2$ was homogeneous in the film: $c(x, t=0) = c_0$. Now, it is safe to assume that the concentration on the surface is constantly zero, $c(0)=c(L)=0$, as the nitrogen flow removes any outgassed molecules. The diffusion equation can be analytically solved for this case (homogeneously filled infinite plan-parallel film) with the given boundary conditions. The solution can be obtained via the method of separation of variables and then using a Fourier series expansion; details of the derivation can be found in the Supplementary Information and in Ref. \onlinecite{BalluffiBook}. The resulting infinite series reads
\begin{equation}\label{eq3}
\begin{split}
    c(x,t)=\frac{4 c_0}{\pi}\sum\limits_{j=0}^{\infty}\left\{\frac{1}{2j+1}\sin\left[(2j+1)\pi\frac{x}{L} \right]\times\right. \\ \left.\exp\left[-\left( \frac{(2j+1)\pi}{L} \right)^2 D t \right] \right\}.
\end{split}
\end{equation}
The ESR intensity is assumed to be a linear function of the concentration integrated for the whole bulk with an added background ($I_0$). Because the concentration changes considerably only through the thickness of the film, the integral needs to be taken only for the thickness, and then it needs to be multiplied by the cross-section of the film. The resulting formula for the intensity is
\begin{widetext}
\begin{equation}
    I_{\text{ESR}}(t) = \tilde{\eta} A \int\limits_0^L c(x,t)~\mathrm{d}x + I_0 = \tilde{\eta} A \frac{8L c_0}{\pi^2}\sum\limits_{j=0}^{\infty}\left\{\frac{1}{(2j+1)^2} \times \exp\left[-\left( \frac{(2j+1)\pi}{L} \right)^2 D t \right] \right\} + I_0.
\end{equation}
\end{widetext}
Here, $\tilde{\eta}$ and $I_0$ are instrumental parameters. Keeping only the first two terms of the series ($j=0,1$), the expression already satisfactorily fits the obtained intensities as presented in Fig. \ref{Fig3:diffusion}b. The diffusion coefficient obtained from the measurement of ClO$_2$ in PET is $D=(3.91 \pm 0.74)\times 10^{-15}$~m$^2$/s at room temperature.

Beyond providing microscopic insight into radical dynamics in PET, the present approach has direct implications for environmental polymer tracking. Spin labeling with ClO$_2$ enables the detection of plastics using ESR spectroscopy, which is inherently insensitive to optical opacity, fluorescence background, or complex chemical matrices. This makes the method particularly attractive for environmental samples such as drinking water, sediments, soils, or wastewater sludges, where conventional optical and thermal techniques often fail. Moreover, the sensitivity of ESR spectral parameters to the local polymer environment suggests that different polymer classes may yield distinguishable spectral signatures, enabling polymer-specific identification.

\section{Conclusions}

In conclusion, we have demonstrated that chlorine dioxide could serve as an efficient, stable, and inorganic spin label for polyethylene terephthalate (PET). Given the similarity in chemical properties, ClO$_2$ might be used for other types of polymers as well, aiding further studies. By exposing PET to ClO$_2$, radicals become incorporated into the polymer matrix, where they remain ESR-active over a wide temperature range and for extended periods of time. Since PET (and most plastics in general) have low microwave absorption, there is no practical limit to the thickness in labeling efficiency. Only the finite time required by the diffusion of the ClO$_2$ molecules has to be considered to achieve a homogeneous distribution in the plastic. On the other hand, there is a technical limitation by the instrumentation in which the sample has to be placed inside a microwave resonator that has a finite active volume of about a few cm$^3$, typically. Temperature-dependent ESR spectroscopy revealed restricted rotational dynamics indicative of hindered molecular motion within the voids of the polymer. The recorded spectra were accurately described by a fitted spin Hamiltonian, reflecting the nature of the environment in the radical's intrinsic electronic properties. Furthermore, time-dependent measurements enabled the determination of the diffusion coefficient of the molecules in the plastic.

These findings establish a simple and general approach to spin labeling otherwise inert plastics without covalent modification or external additives during the manufacturing process. Beyond their use as local probes, embedded ClO$_2$ radicals could provide a platform for the identification, tracing, or degradation monitoring of polymeric waste, polymer-based currencies, and printed circuit boards (PCBs). From an environmental perspective, the demonstrated ClO$_2$-based spin labeling provides a complementary route to existing micro- and nanoplastic detection techniques \cite{Gao2017ASS, Woidasky2020RCR, Grisi2024M, Moon2024SciAdv, Xie2025ACSSRM, Zhang2026CRPS}. Unlike infrared or Raman-based methods, ESR detection does not rely on optical properties and can operate in chemically and structurally complex environments. The proportionality between ESR signal intensity and radical concentration further enables quantitative assessment of polymer content. Time-dependent measurements offer access to diffusion and release kinetics that are relevant for studying polymer aging, transport, and degradation. As such, this approach has the potential to contribute to future strategies for tracing, quantifying, and differentiating polymeric materials in environmental samples, addressing several key challenges highlighted in the Introduction.

The presented method thus opens new perspectives for both fundamental studies of polymer physics and practical applications in sustainable materials research and anti-counterfeiting.

\section*{Acknowledgements}
Work supported by the National Research, Development and Innovation Office of Hungary (NKFIH), and by the Ministry of Culture and Innovation Grants Nr. 149457, 2022-2.1.1-NL-2022-00004, TKP2021-NVA-02.

\section*{Supporting Information}
The Supplementary Information contains the derivation of the time- and space-dependent concentration in the PET film described by Eq. \eqref{eq3}.

\def\bibsection{}
\section*{References}
\bibliographystyle{achemso}
\bibliography{clo2b}

\clearpage
\appendix
\renewcommand{\appendixname}{S}
\renewcommand{\thesection}{S}
\renewcommand\thefigure{\thesection\arabic{figure}}
\setcounter{figure}{0}
\renewcommand\thetable{\thesection\arabic{table}}
\setcounter{table}{0}
\renewcommand{\theequation}{S\arabic{equation}}
\setcounter{equation}{0}
\renewcommand*{\thepage}{S\arabic{page}}
\setcounter{page}{1}

\section*{Supporting Information}

This Supplementary Information contains the derivation of the time- and space-dependent concentration in the PET film described by Eq. (3) of the main text.

\subsection*{Derivation of the Time and Space Dependent Concentration of a Compound Diffusing out of an Infinite Plan-Parallel Medium}

We assume that our compound is nonreactive and that its transport is governed solely by diffusion. Thus, the time evolution of its concentration is described by the diffusion equation:
\begin{equation} \label{diff3d}
    \frac{\partial c(\mathbfit{r}, t)}{\partial t} = D\nabla^2 c(\mathbfit{r},t),
\end{equation}
where $t$ denotes time, $\mathbfit{r}$ the space vector, $c(\mathbfit{r},t)$ the concentration of the compound, $D$ the diffusion coefficient. The medium is assumed to be homogeneous; therefore, $D$ is constant and independent of space and time.

We consider the medium from which $c$ diffuses out to be an infinite plan-parallel volume. Outside the medium, a well-mixed gas phase is assumed. Therefore, the concentration is $0$ there. The bounding surfaces of the medium are taken to be the planes at $x=0$ and $x=L$. Furthermore, we assume that at the initial time ($t=0$) the concentration is spatially uniform within the medium and equal to $c_0$ for $0<x<L$.

In such an arrangement $c$ does not depend on $y$ or $z$, thus Eq. \eqref{diff3d} reduces to a one-dimensional diffusion equation:
\begin{equation} \label{diff1d}
    \frac{\partial c(x, t)}{\partial t} = D \frac{\partial^2 c(x, t)}{\partial x^2}.
\end{equation}
The initial condition is therefore $c(x,0)=c_0$ for $0<x<L$, and $c(x,0)=0$ elsewhere. The boundary conditions are $c(0,t)=c(L,t)=0$.

The solution can be obtained via the separation of variables method. Meaning that we try to find a solution where $c$ is the product of a space-dependent (and time-independent) function and a time-dependent (and space-independent) function, so that 
\begin{equation} \label{prod}
    c(x,t)=X(x) \cdot T(t).
\end{equation}

By substituting Eq. \eqref{prod} into Eq. \eqref{diff1d} and dividing both sides of the equation by $c(x,t)$ we obtain
\begin{equation}
    \frac{\mathrm{d}T(t)/\mathrm{d}t}{T(t)}=D \frac{\mathrm{d}^2 X(x)/\mathrm{d}x^2}{X(x)}.
\end{equation}
Here, the left side depends solely on time, the right side solely on $x$. The equation can hold for arbitrary time and $x$ values only if both sides are equal with the same constant, which we denote by $K$:
\begin{equation}
    \frac{\mathrm{d}T(t)/\mathrm{d}t}{T(t)}=D \frac{\mathrm{d}^2 X(x)/\mathrm{d}x^2}{X(x)}=K.
\end{equation}

This way, we have two ordinary differential equations to solve. We start with the time-dependent part; it can be reformulated as:
\begin{equation}
    \frac{\mathrm{d}T(t)}{\mathrm{d}t}=K T(t),
\end{equation}
which is a linear differential equation. Its solution is $T(t)=\exp(Kt)$, where the multiplicative integration constant is omitted at this stage and will be discussed later.

As the concentration cannot grow infinitely, the constant $K$ must be negative. We therefore introduce the so-called time constant (or characteristic time): $\tau=-1/K$. With them, the solution for the time-dependent part is formulated as
\begin{equation}
    T(t)=\mathrm{e}^{-t/\tau},
\end{equation}
where $\tau>0$ must hold.

Next, we look at the space-dependent function. Our differential equation to solve is the following:
\begin{equation} \label{odex}
    \frac{\mathrm{d}^2X(x)}{\mathrm{d}x^2}=-\frac{1}{\tau D}X(x).
\end{equation}
Furthermore, $X(x)$ must satisfy the same boundary conditions as $c(x, t)$, that is $X(0)=X(L)=0$.

The general solution has the form $X(x)=A\sin(kx)+B\cos(kx)$. Due to the boundary condition of $X(0)=0$, $B$ must be zero. From the other boundary condition, it follows that $kL=n\pi$ must hold, where $n$ is an integer. Furthermore, due to the symmetry of the problem, the solution must remain invariant upon exchanging the two boundaries. This excludes even multiples of $\pi$, and therefore $kL=(2j+1)\pi$ must hold, where $j$ is an integer. The allowed values for $k$ are
\begin{equation} \label{kj}
    k_j=\frac{(2j+1)\pi}{L}.
\end{equation}
Substituting these expressions into Eq. \eqref{odex} yields
\begin{equation}
    -k_j^2 A_j \sin(k_j x)=-\frac{1}{\tau_j D}A_j \sin(k_j x),
\end{equation}
from which the corresponding time constants can be deduced:
\begin{equation} \label{tauj}
    \tau_j=\frac{1}{k_j^2 D}=\left( \frac{L}{(2j+1)\pi} \right)^2 \frac{1}{D}
\end{equation}

The diffusion equation gives infinitely many solutions, and the complete solution for $c$ is given by their linear combination, which is an infinite sum in the following form:
\begin{equation} \label{solution1}
    c(x,t)=\sum\limits_{j=0}^{\infty} A_j \sin(k_j x)\,\mathrm{e}^{-k_j^2 Dt}.
\end{equation}
This sum should contain linearly independent terms; therefore, negative values for $j$ were excluded (due to symmetry reasons, $\sin(k_{-j}x)=-\sin(k_{j-1}x)$).

The initial condition can be used to determine the coefficients $A_j$. Setting $t=0$ in Eq. \eqref{solution1} yields
\begin{equation}
    c(x,0)=\sum\limits_{j=0}^{\infty} A_j \sin(k_j x) =c_0.
\end{equation}
This shows that the coefficients, $A_j$, are the Fourier-coefficients of the initial value.

To calculate these, both sides are multiplied by $\sin(k_i x)$ and integrated over the domain:
\begin{equation}
    \int\limits_0^L \left\{ \sin(k_i x)\sum\limits_{j=0}^{\infty} A_j \sin(k_j x)  \right\} \,\mathrm{d}x=\int\limits_0^L c_0 \sin(k_i x) \,\mathrm{d}x.
\end{equation}
Interchanging the order of summation and integration (as both $A_j$ and $k_j$ are independent of $x$) leads to
\begin{equation} \label{Fourier}
    \sum\limits_{j=0}^{\infty}\left\{ A_j \int\limits_0^L \sin(k_j x)\sin(k_i x) \,\mathrm{d}x \right\}=c_0\int\limits_0^L \sin(k_i x) \,\mathrm{d}x.
\end{equation}

We remark that due to Eq. \eqref{kj}, $\sin(k_j L)=0$, and $\cos(k_j L)=-1$. This allows the right-hand side to be evaluated as:
\begin{equation}
    c_0\int\limits_0^L \sin(k_i x) \,\mathrm{d}x = - c_0 \left. \frac{\cos(k_i x)}{k_i} \right|_0^L=\frac{2c_0}{k_i}.
\end{equation}

By using the product-to-sum trigonometric identity, the integral on the left-hand side can be reformulated as follows:
\begin{align}
    &\int\limits_0^L \sin(k_j x)\sin(k_i x) \,\mathrm{d}x = \nonumber \\  =~&\frac{1}{2}\int\limits_0^L \cos[(k_j-k_i)x]\,\mathrm{d}x-\frac{1}{2}\int\limits_0^L \cos[(k_j+k_i)x]\,\mathrm{d}x.
\end{align}
For $i \neq j$, both $(k_j-k_i)L$ and $(k_j+k_i)L$ are even multiples of $\pi$, thus both integrals are taken over whole periods, meaning they are $0$. In the case of $i=j$, this is only true for the second integral, $\cos[(k_i-k_i)x]=1$, thus the first term evaluates as $L/2$. Substituting all of these integral values to Eq. \eqref{Fourier} yields
\begin{equation}
    A_i \frac{L}{2}=\frac{2c_0}{k_i}.
\end{equation}
Reordering for $A_i$, relabeling the index $i$ to $j$, and using Eq. \eqref{kj} gives
\begin{equation} \label{Aj}
    A_j=\frac{4 c_0}{k_j L}=\frac{4 c_0}{(2j+1)\pi}.
\end{equation}

Finally, substituting Eqs. \eqref{kj} and \eqref{Aj} into \ref{solution1} gives the complete time- and space-dependent solution of Eq. \eqref{diff1d}:
\begin{equation}
\begin{split}
    c(x,t)=\frac{4 c_0}{\pi}\sum\limits_{j=0}^{\infty}\left\{\frac{1}{2j+1}\sin\left[(2j+1)\pi\frac{x}{L} \right]\times\right. \\ \left.\exp\left[-\left( \frac{(2j+1)\pi}{L} \right)^2 D t \right] \right\}.
\end{split}
\end{equation}

\end{document}